\documentclass[journal]{IEEEtran}
\usepackage[applemac]{inputenc}
\usepackage[english]{babel}

\usepackage{algorithm,
	algorithmic,
	amsmath,
	amssymb,
	amsthm,
	thmtools}

\usepackage{cite}

\usepackage{xcolor,
	graphicx,
	tikz,
	pgfplots}
	
\graphicspath{{Figures/}}
	
\pgfplotsset{compat=1.5}
\colorlet{color1}{red}
\colorlet{color2}{green!60!black}
\colorlet{color3}{blue}
\colorlet{color4}{magenta}

\usetikzlibrary{calc,
shapes,
positioning,
patterns,
decorations.pathreplacing}
	
\colorlet{green}{green!60!black}

\makeatletter
\tikzset{% customization of pattern
        hatch distance/.store in=\hatchdistance,
        hatch distance=5pt,
        hatch thickness/.store in=\hatchthickness,
        hatch thickness=5pt
        }
\pgfdeclarepatternformonly[\hatchdistance,\hatchthickness]{north east hatch}% name
    {\pgfqpoint{-1pt}{-1pt}}% below left
    {\pgfqpoint{\hatchdistance}{\hatchdistance}}% above right
    {\pgfpoint{\hatchdistance-1pt}{\hatchdistance-1pt}}%
    {
        \pgfsetcolor{\tikz@pattern@color}
        \pgfsetlinewidth{\hatchthickness}
        \pgfpathmoveto{\pgfqpoint{0pt}{0pt}}
        \pgfpathlineto{\pgfqpoint{\hatchdistance}{\hatchdistance}}
        \pgfusepath{stroke}
    }
\makeatother

\usepackage{multirow}

\usepackage{dblfloatfix}
\newcounter{mytempeqncnt}

\IEEEoverridecommandlockouts

\newtheorem{proposition}{Proposition}
\newtheorem{theorem}{Theorem}
\newcommand{\E}{\mathbb{E}}
\renewcommand{\Pr}{\mathbb{P}}

\newcommand{\0}{\boldsymbol{0}}
\newcommand{\1}{\boldsymbol{1}}
\newcommand{\ba}{\boldsymbol{\alpha}}
\newcommand{\apop}{\ba_\textnormal{pop}}
\newcommand{\arlx}{\check{\ba}^*}
\newcommand{\afloor}{\hat{\ba}^*}
\newcommand{\anc}{\ba_\textnormal{nc}}
\newcommand{\astr}{\ba'}

\newcommand{\calA}{\mathcal{A}}
\newcommand{\calC}{\mathcal{C}}
\newcommand{\calCbar}{\bar{\mathcal{C}}}

\newcommand{\checka}{\check{\alpha}_i^*}

\newcommand{\bbar}{\beta_\textnormal{d}}
\newcommand{\bfloor}{{\lfloor\beta\rfloor}}
\newcommand{\gBSC}{\gamma_{\textnormal{BS}}^{(\calC)}(i,j)}
\newcommand{\gBSCbar}{\gamma_{\textnormal{BS}}^{(\calCbar)}(i,j)}
\newcommand{\gDC}{\gamma_{\textnormal{D2D}}^{(\calC)}(i,j)}
\newcommand{\gDCbar}{\gamma_{\textnormal{D2D}}^{(\calCbar)}(i,j)}
\newcommand{\smin}{s_\textnormal{min}}
\newcommand{\smax}{s_\textnormal{max}}
\newcommand{\Ta}{T_\textnormal{a}}
\newcommand{\Tc}{T_\textnormal{c}}
\newcommand{\Ti}{T_\textnormal{i}}
\title{Optimizing MDS Coded Caching in Wireless Networks with Device-to-Device Communication}

\begin{document}

\author{Jesper Pedersen, Alexandre Graell i Amat,~\IEEEmembership{Senior Member,~IEEE},\\ Iryna Andriyanova,~\IEEEmembership{Member,~IEEE}, and Fredrik Br\"annstr\"om,~\IEEEmembership{Member,~IEEE}
\thanks{This work was funded by the Swedish Research Council under grant 2016-04253 and by the National Center for Scientific Research in France under grant CNRS-PICS-2016-DISCO.}
\thanks{J. Pedersen, A. Graell i Amat, and F. Br\"annstr\"om are with the Department of Electrical Engineering, Chalmers University of Technology, SE-41296 Gothenburg, Sweden (e-mail: \{jesper.pedersen, alexandre.graell, fredrik.brannstrom\}@chalmers.se).}
\thanks{I. Andriyanova is with the ETIS-UMR8051 group, ENSEA/University of Cergy-Pontoise/CNRS, 95015 Cergy, France (e-mail: iryna.andriyanova@ensea.fr).}
\vspace{-3ex}}

\maketitle

\begin{abstract}
We consider the caching of content in the mobile devices in a dense wireless network using maximum distance separable (MDS) codes. We focus on an area, served by a base station (BS), where mobile devices move around according to a random mobility model. Users requesting a particular file download coded packets from caching devices within a communication range, using device-to-device communication. If additional packets are required to decode the file, these are downloaded from the BS. We analyze the device mobility and derive a good approximation of the distribution of caching devices within the communication range of mobile devices at any given time. We then optimize the MDS codes to minimize the network load under a cache size constraint and show that using optimized MDS codes results in significantly lower network load compared to when caching the most popular files. We further show numerically that caching coded packets of each file on all mobile devices, i.e., maximal spreading, is optimal.
\end{abstract}

\begin{IEEEkeywords}
	Caching, content delivery, device-to-device communication, device mobility, erasure correcting codes.
\end{IEEEkeywords}

\section{Introduction}
Mobile data traffic is predicted to increase significantly in the coming years \cite{Ericsson2017}, which imposes a severe strain on existing wireless networks. One of the most promising methods to offload traffic is storing content closer to the end users, a technique known as \emph{caching} \cite{Maddah-Ali2014, Andrews2014, Liu2016, Boccardi2014, Shanmugam2013}. Content can be cached at small base stations (BSs) to reduce the burden on the backhaul links \cite{Andrews2014, Liu2016}. Alternatively, content may be cached directly in the mobile devices, which helps in reducing both downlink traffic from BSs \cite{Liu2016, Boccardi2014} and the backhaul traffic. %It has been shown that caching content using erasure correcting codes can significantly improve the performance in wireless networks when a user requesting content can access only a subset of the caches \cite{Shanmugam2013,Bioglio2015,Ji2016,Pedersen2015, Pedersen2016, Piemontese2016}.}

A plethora of works on coded caching has appeared in recent years. In the literature, the concept of coded caching refers to both the caching of uncoded content to facilitate index-coded broadcasts \cite{Maddah-Ali2014, Ji2016}, and the use of erasure correcting codes to cache the content \cite{Shanmugam2013,Bioglio2015,Ji2016,Golrezaei2014,Pedersen2015, Pedersen2016, Piemontese2016, Piemontese2018, Paakkonen2013, Wang2017}. In both cases, the goal is to deliver content efficiently. In \cite{Maddah-Ali2014}, index coding is shown to significantly reduce the amount of data that is required to transmit over a shared link. In \cite{Ji2016}, content is cached directly in the mobile devices. Asymptotic scaling laws of the amount of data necessary to transmit to satisfy worst case file demands using index-coded device-to-device (D2D) broadcasts for fixed network topologies is investigated for the case where the file size, the number of files in the library, and the number of users grows large. An additional layer of erasure correcting codes is suggested to facilitate a decentralized caching scheme.

Erasure correcting codes can significantly improve the performance in wireless networks when a user requesting content can access only a subset of the caches \cite{Shanmugam2013,Bioglio2015,Golrezaei2014,Pedersen2015, Pedersen2016, Piemontese2016, Piemontese2018, Paakkonen2013, Wang2017}. In \cite{Shanmugam2013} and \cite{Bioglio2015}, files are cached in a number of small BSs from which mobile devices download content. It is shown that caching content using maximum distance separable (MDS) codes reduces the download delay and that the performance improves with an increase in the density of small BSs and a decrease of the density of devices in the network \cite{Shanmugam2013}. In \cite{Bioglio2015}, the use of MDS codes is shown to reduce the amount of data that is required to download from the macro BS.  In \cite{Golrezaei2014, Paakkonen2013, Pedersen2015, Pedersen2016, Piemontese2016, Piemontese2018,  Wang2017} caching coded content directly in the mobile devices is considered and devices download requested files using D2D communication. 

The caching of complete files in the mobile devices, where the devices move around according a simple random walk model, is considered in \cite{Golrezaei2014}. Files are cached randomly according to a Zipf distribution and the Zipf parameter is optimized to maximize the number of times a requested file can be found in the cache of a nearby device. Coded caching in the mobile devices considering device mobility has been studied in \cite{Paakkonen2013, Pedersen2015, Pedersen2016, Piemontese2016,Piemontese2018,Wang2017}. In \cite{Paakkonen2013, Pedersen2015, Pedersen2016}, devices arrive to and depart from an area according to a Poisson process and coding is shown to reduce the amount of data required to download requested files. In \cite{Piemontese2016,Piemontese2018}, the use of MDS codes is shown to minimize the download delay. All these previous works  \cite{Paakkonen2013, Golrezaei2014, Pedersen2015, Pedersen2016, Piemontese2016} assume an area-centric model, where all devices within an area can communicate with each other, regardless of the distance between them. A more realistic model is a user-centric model where a device can communicate only with neighboring devices within a given communication range. To the best of our knowledge,  \cite{Wang2017} is the only paper that considers a user-centric mobility model and studies the effects of coded caching with device mobility. However, \cite{Wang2017} assumes that the devices remain within the communication range for a deterministic time and that a file can be reconstructed from a random number of coded packets independent of the content allocation. Also, \cite{Wang2017} considers only small networks in terms of number of users, corresponding to low device densities.

 %The various performance gains of caching MDS coded packets, as opposed to only cache copies of the most popular files, i.e., using a repetition code, observed in \cite{Bioglio2015, Shanmugam2013, Pedersen2015, Pedersen2016, Wang2017} motivates further research of caching MDS coded packets at the wireless edge.

\subsection{Contribution}

In this paper, we study the effect of MDS-coded caching of content in the mobile devices to reduce the network load (from the BS and the mobile devices) in highly-dense wireless networks considering device mobility. As in \cite{Wang2017}, we consider a user-centric model, and assume that the devices move around an area according to the random waypoint model \cite{Johnson1996} and request files from a library at random times. Files are encoded using MDS codes of equal code length but potentially different code rate, and coded packets are cached in a number of mobile devices. When a device requests a file from the library, coded packets are downloaded from mobile devices within the communication range using D2D communication, and if additional packets are required to decode the requested file, these are retrieved from the BS. We analyze the mobility model and derive a good approximation of the distribution of the number of devices within range at the time of a request. We then formulate the minimization of the network load as a mixed integer linear program (MILP) that allows us to find the content allocation that minimizes the network load, i.e., minimizes the amount of data that is downloaded from the BS and mobile devices, assuming a global average cache size constraint (across all devices). We also suggest a greedy algorithm to enforce a strict cache size constraint per device. The problem formulation includes a weighting parameter to reflect the cost of utilizing the downlink and D2D communication. For a number of devices up to $\sim1000$, we can solve the MILP using a branch-and-bound method that guarantees that the global minimum is attained. For a larger number of devices, i.e., higher device density scenarios, we propose a relaxation of the integer constraint of the MILP into a linear program (LP) which provides a lower bound on the network load. We also give a simple suboptimal algorithm to find an upper bound on the network load. We show numerically that caching packets of a given file on all mobile devices, i.e., maximal spreading \cite{Leong2012}, is optimal. We further show numerically for maximal spreading that the upper and lower bounds are approximately equal. Hence, the proposed lower and upper bounds provide a very good approximation to the optimal solution. Compared to \cite{Wang2017}, our formulation allows to analyze highly-dense networks.

%{\color{magenta}The work closest to ours is \cite{Wang2017}, where the main limitation is that relatively small networks are analyzed. This paper addresses several important issues that distinguishes from \cite{Wang2017}. Specifically:
%\begin{itemize}
%	\item We investigate the network load gains for dense networks and show that optimizing the rates of the MDS codes may yield a significantly lower network load than when caching only the most popular files.
%\end{itemize}}

\section{System Model}\label{sec:model}
We consider a cell with area $A$ square meters served by a BS. We assume a higher level inter-cell interference coordination \cite{Kosta2013} such that we can consider inter-cell interference to be negligible, which enables us to analyze one cell in isolation. The area is projected onto a sphere of radius $\rho$ meters to remove the area boundaries \cite{Gupta2000}, where $A = 4\pi\rho^2$. $M$ mobile devices are uniformly spread over the area. Users wish to download files from a library of $N$ files that is always accessible to the BS\footnote{This assumption is valid if we consider the BS to be connected to the core internet through a high capacity optical fiber backhaul link.}. We assume that all files have equal size $F$ bits, which is without loss of generality since contents can always be divided into chunks with equal size \cite{Li2018}. Similar to most previous works on coded caching, see, e.g., \cite{Bioglio2015, Ji2016:jsac}, we assume that the file popularity follows the Zipf distribution \cite{Breslau1999}. Hence, the popularity of file $i$ is given by
\begin{equation}\label{eq:zipf}
	p_i = \frac{1/i^\sigma}{\sum_{\ell=1}^N 1/\ell^\sigma},~ i = 1, \ldots, N,~ 0 < \sigma \le 1.5,
\end{equation}
where $\sigma$ is the skewness parameter of the distribution. Note that, although all our results are obtained assuming a Zipf distribution of file popularities, the framework is general in the sense that other distributions, such as the Weibull and Gamma distributions that are suggested alternatives for YouTube videos in \cite{Cheng2008}, can be used.

\subsection{Content Allocation}
\begin{figure}
	\centering
	\begin{tikzpicture}[scale = 1,
		>=stealth,
		Pattern/.style={pattern=north east hatch, pattern color=white, hatch distance=5pt, hatch thickness=1.3pt}
		]
		\tikzstyle{content}=[shape=rectangle,
			rounded corners=1pt,
			inner sep=0pt,
			minimum height=10pt,
		]
				
		\node[content, draw=red, fill=red, minimum width=60pt, anchor=west] at (0,0) {};
		\node[content, draw=green, fill=green, minimum width=60pt, anchor=west] at (0,-0.75) {};
		
		\draw[->] (2.5,-0.1) -- (3.5,-0.1) node[above, midway] {\footnotesize $(n,k_1)$};
		\draw[->] (2.5,-0.85) -- (3.5,-0.85) node[above, midway] {\footnotesize $(n,k_2)$};
		
		\foreach \i in {0,1}{
			\node[content, draw=red, fill=red, anchor=west, minimum width=30pt] at (1.2*\i+3.8,0) {};
		}
		
		\foreach \i in {0,1,2}{
			\node[content, draw=green, fill=green, anchor=west, minimum width=20pt] (info\i) at (0.85*\i+3.8,-0.75) {};
		}
		
		% Coded symbols
		\foreach \i in {0,1}{
			\node[content, preaction={fill=red}, Pattern, draw=red, inner sep=0pt, minimum width=30pt, anchor=west] (coded\i) at (1.2*\i+6.2,0) {};
		}
		
		\node[content, preaction={fill=green}, Pattern, draw=green, inner sep=0pt, minimum width=20pt, anchor=west] at (6.35,-0.75) {};
		
		\draw[decorate, decoration={brace, mirror, amplitude=3pt}, xshift=-1em, yshift=-2em] ($(coded1.south west)-(0,2pt)$) -- ($(coded1.south east)-(0,2pt)$) node [black, midway, yshift=-1em]{\footnotesize $\alpha_1$};
		\draw[decorate, decoration={brace, mirror, amplitude=3pt}, xshift=-1em, yshift=-2em] ($(info1.south west)-(0,2pt)$) -- ($(info1.south east)-(0,2pt)$) node [black, midway, yshift=-1em]{\footnotesize $\alpha_2$};
	
	\end{tikzpicture}
	\vspace{-3ex}
	\caption{Encoding example for the caching of two files (red and green). In the example, $n=4$, $k_1=2$, and $k_2=3$. The solid rectangles (to the right) represent the $k_i$ packets which together with the dashed rectangles represent the $n$ coded packets.}\label{fig:enc}
	\vspace{-3ex}
\end{figure}
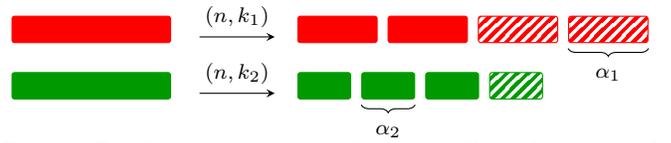

Each file $i$ that is to be cached is partitioned into $k_i$ packets, each of size $F/k_i$ bits, and encoded into $n$ packets, also of size $F/k_i$ bits, using an $(n, k_i)$ MDS code of code length $n$, dimension $k_i=1,\ldots,n$, and rate $R_i=k_i/n\le1$ \cite{Ryan2009}. Thus, different files are encoded by MDS codes of the same code length but potentially of different dimension, i.e., different rate. The $n$ coded packets are cached in $n$ mobile devices (possibly different for each file) in the area, chosen uniformly at random. Hence, for each file $i$, each of the $n$ devices caching the file stores one coded packet of the file, i.e., a fraction $\alpha_i=1/k_i$ of the file. Thus, as $k_i=1,\ldots,n$,
\begin{equation}
\label{eq:alpha}
	\alpha_i \in \left\{0, 1/n, 1/(n-1), \ldots, 1\right\} \triangleq \calA,~ i = 1, \ldots, N,
\end{equation}
where $\alpha_i = 0$ implies that file $i$ is not cached. A small illustrative example where two files are to be cached using codes of parameters $(n,k_1)$ and $(n,k_2)$, with $n=4$, $k_1=2$, and $k_2=3$, is shown in Fig.~\ref{fig:enc}. We define the vector $\ba = (\alpha_1, \ldots, \alpha_N)$ and refer to it as the \emph{content allocation}. Note that the content allocation is inversely proportional to the code rate as
\begin{equation}
	R_i = \frac{1}{n\alpha_i}.
\end{equation}

In practice, a strict cache size constraint per device would be desirable. Unfortunately, this leads to a very complicated optimization problem. To simplify the problem and similar to \cite{Leong2012}, we enforce a global average cache size constraint, denoted by $\beta$, where
\begin{equation}\label{eq:beta}
	\sum_{i=1}^N \alpha_i \le \beta.
\end{equation}
This implies an average cache size constraint per device
\begin{equation}\label{eq:bbar}
	\bbar = \beta n/M.
\end{equation}
In Section~\ref{sec:min}, we suggest a suboptimal greedy algorithm that enforces a strict cache size constraint per device and show numerically in Section~\ref{sec:results} that the incurred performance loss is negligible for a small cache size overhead. We remark that for $n=M$, i.e., maximal spreading, the average cache size constraint becomes a strict cache size constraint.

The commonly used \emph{popular} content allocation, where each of the $\bfloor$ most popular files is cached in $n$ (possibly different) mobile devices (i.e., using an $(n,1)$ repetition code), is given by
\begin{equation}\label{eq:apop}
	\apop = \left(\1_\bfloor, \0_{N-\bfloor}\right),
\end{equation}
where $\1_\bfloor$ is a vector with $\bfloor$ ones and $\0_{N-\bfloor}$ is a vector with $N-\bfloor$ zeros.

\subsection{Data Download}
Mobile devices request files at random times, with the time between requests exponentially independent, identically distributed (i.i.d.) with rate $\omega$ per second, i.e., the request process is a Poisson process \cite{Miller2004}. Hence, the expected total request rate in the area is $M\omega$. A device requests file $i$ with probability $p_i$ given by \eqref{eq:zipf}. Due to the MDS property, $k_i$ coded packets are sufficient to decode the file \cite{Ryan2009}. The user requesting content downloads as many coded packets as possible (up to $k_i$) from caching devices within a communication radius of $r$ meters (measured over the curvature of the sphere), referred to as the communication range. If additional packets are required, these are retrieved from the BS. The equivalent number of files that are downloaded from the BS per second is referred to as the \emph{downlink rate} and the equivalent number of files downloaded from caching devices (per second) is referred to as the \emph{D2D communication rate}. We assume that D2D interference can be considered negligible, which can be achieved by considering the coordination of the radio resources using, e.g., a scheme similar to the one suggested in \cite{Sun2016}. Similar assumptions are made in \cite{Paakkonen2013} and \cite{Wang2017}. We furthermore assume that the communication is error free and incurs zero delay.

\subsection{Device Mobility}
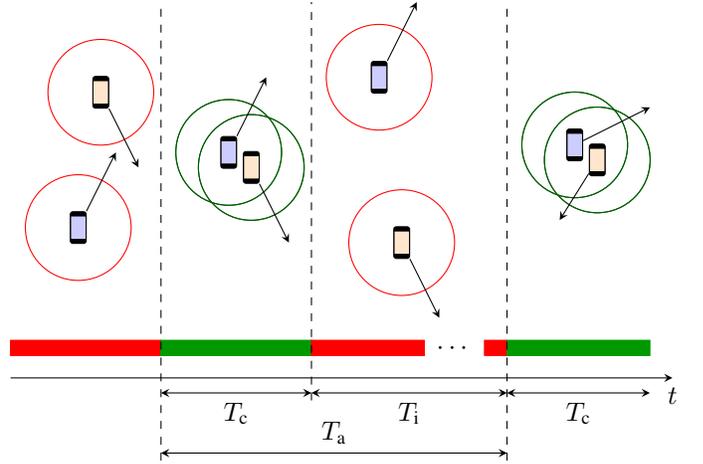
\begin{figure}
	\centering
	\begin{tikzpicture}[>=stealth,
		scale = 1
	]
		\tikzstyle{device} = [shape = rectangle,
			draw = black,
			fill = black,
			rounded corners = 1pt,
			inner sep = 0pt,
			minimum width=6pt,
			minimum height = 12pt
		]
		\tikzstyle{screen1} = [shape = rectangle,
			draw = blue!20,
			fill = blue!20,
			inner sep = 0pt,
			minimum width = 5pt,
			minimum height = 8pt
		]
		\tikzstyle{screen2} = [shape = rectangle,
			draw = orange!20,
			fill = orange!20,
			inner sep = 0pt,
			minimum width = 5pt,
			minimum height = 8pt
		]
		\draw[->] (0,-3) to (8.8,-3) node[below] {$t$};
		\filldraw[red] (0,-2.7) rectangle (2,-2.5);
		\filldraw[green] (2,-2.7) rectangle (4,-2.5);
		\filldraw[red] (4,-2.7) rectangle (5.5,-2.5);
		\node at (5.9,-2.6) {$\cdots$};
		\filldraw[red] (6.3,-2.7) rectangle (6.6,-2.5);
		\filldraw[green] (6.6,-2.7) rectangle (8.5,-2.5);
		
		\draw[dashed] (2,-4.1) to (2,2);
		\draw[dashed] (4,-3.3) to (4,2);
		\draw[dashed] (6.6,-4.1) to (6.6,2);
		
		\draw[<->] (2,-3.2) to node[below, midway] {$\Tc$} (4,-3.2);
		\draw[<->] (4,-3.2) to node[below, midway] {$\Ti$} (6.6,-3.2);
		\draw[<->] (2,-4) to node[above, midway] {$\Ta$} (6.6,-4);
		\draw[<->] (6.6,-3.2) to node[below, midway] {$\Tc$} (8.5,-3.2);

		\foreach \j in {1,2,3}
			{
				\node[device] at ($(2*\j-1,0)+(0,\j-2)+(-0.1,0)$) (d1\j) {};
				\node[screen1] at (d1\j.base) {};
				\node[device] at ($(2*\j-1,0.5)+(0,1.5-\j)+(0.2,-0.2)$) (d2\j){};
				\node[screen2] at (d2\j.base) {};
			}

%		\node at (6.2,0) {$\cdots$};
		
		\node[device] at (7.5,0.1) (d14) {};
		\node[screen1] at (d14.base) {};
		\node[device] at (7.8,-0.1) (d24) {};
		\node[screen2] at (d24.base) {};
		
		\foreach \i in {1,2}
			\foreach \j in {1,3}
			{
				\draw[red] (d\i\j) circle (20pt);
				\draw[green!60!black] (d\i2) circle (20pt);
				\draw[green!60!black] (d\i4) circle (20pt);
			}
		
		\foreach \j in {1,2,3}
			{
				\draw[->] (d1\j) to ($(d1\j)+(0.5,1)$);
				\draw[->] (d2\j) to ($(d2\j)+(0.5,-1)$);
			}
		\draw[->] (d14) to ($(d14)+(1,0.5)$);
		\draw[->] (d24) to ($(d24)-(0.5,0.8)$);
	\end{tikzpicture}
	\vspace{-3ex}
	\caption{The device mobility model with contact, intercontact, and interarrival times illustrated for two devices. In frames 2 and 4, the devices are within each others communication range and in frames 1 and 3, the devices are not within range.}\label{fig:model}
	\vspace{-3ex}
\end{figure}

The mobile devices move around the area according to the random waypoint model \cite{Johnson1996}, which was compared with realistic data sets for a smaller number of mobile devices in \cite{Wang2017}. According to this model, a mobile device pauses for a deterministic time and then picks a target uniformly in the area and a speed uniformly between a minimum and a maximum speed, denoted by $\smin$ and $\smax$, respectively. For simplicity, we assume that the pause time is zero but the analysis and results are easily generalized to account for a nonzero pause time using the results in \cite{Abdulla2007}. The device traverses the great circle towards the target and, once the target has been reached, repeats the process. The targets and speeds are i.i.d. for all devices in the area. As two mobile devices move around the area, they are within the communication range for a random \emph{contact time}, denoted by $\Tc$ and measured in seconds. The time between two contacts is referred to as the \emph{intercontact time}, denoted by $\Ti$, and the time between the beginning of two contacts is referred to as the \emph{interarrival time}, denoted by $\Ta$, where
$$
	\Ta = \Tc + \Ti.
$$
The device mobility model is illustrated in Fig.~\ref{fig:model}.

\section{Network Statistics Analysis}
In this section, we derive the probability to have a number of caching devices within the communication range at the time of a request, as well as the amount of data that is downloaded to serve user requests. We have the following theorem.

\begin{theorem}\label{thm:qj}
	Consider the scenario in Section~\ref{sec:model} with an area $4\pi\rho^2$, $M$ mobile devices with communication range $r$, a minimum and maximum speed of devices $\smin$ and $\smax$, respectively, and a code length $n$. Under the assumption that $r \ll 2\rho$ and $\smin \approx \smax$, the probability that there are $j$ caching devices within the communication range of any device is
	\begin{equation}\label{eq:qj}
		q_j = \frac{\left(\frac{\lambda}{\mu} \cdot \frac{n}{M}\right)^j}{j!} \cdot e^{-\frac{\lambda}{\mu} \cdot \frac{n}{M}},\quad j \ge 0,
	\end{equation}
	where the aggregate arrival rate of devices to within the communication range of any mobile device, denoted by $\lambda$, is
%	\begin{equation}\label{eq:lambdamu}
%		\lambda/\mu = (M-1) \left(\frac{r}{2\rho}\right)^2.
%	\end{equation}
	\begin{equation}\label{eq:lambda}
		\lambda = (M-1) \frac{2rs}{4\pi\rho^2}
	\end{equation}
	and the departure rate, denoted by $\mu$, is
	\begin{equation}\label{eq:mu}
		\mu = \frac{2s}{\pi r}.
	\end{equation}
\end{theorem}
\begin{IEEEproof}
For the random waypoint model, under the assumption that $\smin \approx \smax$, two devices move at an approximate average relative speed \cite{Abdulla2007}
$$
	s = \frac{2(\smax+\smin)}{\pi}
$$
and the expected contact time is \cite{Abdulla2007}
\begin{equation}\label{eq:ETc}
	\E[\Tc] = \frac{\pi r}{2s},
\end{equation}
which holds under the assumption that a device does not change direction during the contact time. This assumption is valid when the communication area $\pi r^2$ is much smaller than the area $A$, i.e., when $r \ll 2\rho$. The departure rate is given by
\begin{equation}
	\mu = 1/\E[\Tc],
\end{equation}
which inserted in \eqref{eq:ETc} gives \eqref{eq:mu}.

The distribution of $\Ti$ can be closely approximated by
\begin{equation}\label{eq:Ti}
	\Ti \overset{\text{i.i.d.}}{\sim} \text{Exp}\left(\frac{2rs}{4\pi\rho^2}\right),
\end{equation}
for $r \ll 2\rho$ \cite{Abdulla2007}. The interarrival time is
\begin{equation}\label{eq:Ta}
	\Ta = \Tc + \Ti \approx \Ti,
\end{equation}
since $r \ll 2\rho$ implies $\Tc \ll \Ti$. Using \eqref{eq:Ti} and \eqref{eq:Ta}, the aggregate arrival rate is the sum of the interarrival rates of $M-1$ devices in the area and we get \eqref{eq:lambda} \cite[Sec.~1.3.1.2]{Bolch2006}.

\begin{figure*}[!t]
\normalsize
\setcounter{mytempeqncnt}{\value{equation}}
\setcounter{equation}{25}
\begin{equation}\label{eq:hexpand}
	h(\ba) = M\omega \sum_{i=1}^N p_i \sum_{j=0}^\infty q_j\max \left\{ \alpha_i\left((1-2\theta)j-\theta\frac{n}{M}\right)+\theta, (1-\theta) \left( 1-\alpha_i \frac{n}{M} \right) \right\}
\end{equation}
\hrulefill
%\vspace*{-1ex}
\setcounter{equation}{\value{mytempeqncnt}}
\end{figure*}

To simplify the analysis, we assume that the arrival rate is independent of the number of devices within the communication range, which is a reasonable assumption for $r\ll2\rho$. Under this assumption, the stochastic process describing the number of mobile devices within the communication range of any reference device can be characterized by an $\mathsf{M}/\mathsf{G}/\infty$ queueing model \cite[Sec.~6.1.1]{Bolch2006}. Since the interarrival times are independent and homogeneous, the steady-state distribution of the number of devices within the communication range of any reference device is Poisson with mean \cite{Newell1966}
\begin{equation}
	\int_0^\infty \lambda [1-\Pr(\Tc \le t)]~dt = \lambda \E[\Tc] = \lambda/\mu.
\end{equation}
A fraction $n/M$ of the devices cache a packet of a particular file. Hence, the expected number of caching devices within the communication range is \cite[Sec.~1.3.1.2]{Bolch2006}
$$
	\frac{\lambda}{\mu} \cdot \frac{n}{M},
$$
which gives \eqref{eq:qj} and the proof is complete.
\end{IEEEproof}
For a Poisson point process, $M-1$ devices are uniformly spread over an area $A$. The number of devices within a communication range $r$ follows the Poisson distribution with mean \cite{Haenggi2013}
\begin{equation}\label{eq:ppp}
	(M-1) \frac{\pi r^2}{A}.
\end{equation}
Note that, for such a process, independent realizations of the device locations are assumed. For the device mobility model considered in Section~\ref{sec:model}, the device locations are uniformly distributed over the sphere, but the location of a device at any given time is dependent of the location at the previous time instant. Interestingly, using \eqref{eq:lambda} and \eqref{eq:mu}, \eqref{eq:ppp} is equal to $\lambda/\mu$ and the distribution of devices within the communication range of any reference device is the same according to Theorem~\ref{thm:qj}.

We denote by $\calC_i$ and $\calCbar_i$ the events that the request for file $i$ comes from a device caching or not caching a coded packet of file $i$, respectively, where
\begin{equation}\label{eq:PrC}
	\Pr(\calC_i) = 1-\Pr(\calCbar_i) = \frac{n}{M},\quad i = 1, \ldots, N.
\end{equation}
The amount of data that is downloaded from the BS and from the mobile devices is given by the following proposition.
\begin{proposition}\label{prop:gamma}
	Assume a device requests file $i$ and there are $j$ devices within its communication range caching a coded packet of file $i$. If the device caches a coded packet of file $i$, the fraction of file $i$ that is downloaded from the BS is
	\begin{equation}\label{eq:gBSC}
		\gBSC = \begin{cases}
			1-\alpha_i(j+1), & \text{if $0 \le j < 1/\alpha_i$}\\
			0, & \text{if $j \ge 1/\alpha_i$}
		\end{cases}.
	\end{equation}
	Otherwise, the corresponding fraction of file $i$ that is downloaded from the BS is
	\begin{equation}\label{eq:gBSCbar}
		\gBSCbar = \begin{cases}
			1-j\alpha_i, & \text{if $0 \le j < 1/\alpha_i$}\\
			0, & \text{if $j \ge 1/\alpha_i$}
		\end{cases}.
	\end{equation}
	Moreover, the fraction of file $i$ that is downloaded from $j$ caching devices is
	\begin{equation}\label{eq:gDC}
		\gDC = 1-\alpha_i-\gBSC,
	\end{equation}
	and
	\begin{equation}\label{eq:gDCbar}
		\gDCbar = 1-\gBSCbar
	\end{equation}
	depending on whether the device requesting file $i$ caches a packet of file $i$ or not.
\end{proposition}
\begin{IEEEproof}
	A device requires $k_i$ coded packets to decode a requested file $i$. If there are $j < k_i=1/\alpha_i$ devices caching a packet of file $i$ within the communication range, $k_i-j$ packets are retrieved from the BS if the device placing the request does not cache a packet of file $i$. If the device caches a packet of file $i$, $k_i-j-1$ packets are downloaded from the BS. If $j \ge k_i$, no packets are downloaded from the BS. Hence, the fraction of file $i$ that is downloaded from the BS is given by \eqref{eq:gBSC} and \eqref{eq:gBSCbar}. Following the same argument, one obtains \eqref{eq:gDC} and \eqref{eq:gDCbar}.
\end{IEEEproof}

The expected downlink rate for a content allocation $\ba$, denoted by $f(\ba)$, is given by
\begin{equation}\label{eq:f}
	f(\ba) = M\omega \sum_{i = 1}^N p_i \sum_{j = 0}^\infty q_j\left(\gBSCbar\Pr(\calCbar_i)+\gBSC\Pr(\calC_i)\right),
\end{equation}
where, using \eqref{eq:PrC}--\eqref{eq:gBSCbar},
\begin{align}
	& \gBSCbar \Pr(\calCbar_i) + \gBSC \Pr(\calC_i) \nonumber \\
	& ~~~~~~~~~~~~~ = \begin{cases}
		1-\alpha_i\left(j+\frac{n}{M}\right), & \text{if $0 \le j < 1/\alpha_i$}\\
		0, & \text{if $j \ge 1/\alpha_i$}
	\end{cases}.
\end{align}
The expected D2D communication rate, denoted by $g(\ba)$, is given by
\begin{equation}
\label{eq:g}
	g(\ba) = M\omega \sum_{i = 1}^N\hspace{-1pt} p_i \sum_{j = 0}^\infty q_j \left(\gDCbar \Pr(\calCbar_i)+\gDC \Pr(\calC_i)\right),
\end{equation}
where, using \eqref{eq:PrC}--\eqref{eq:gDCbar},
\begin{align}
	& \gDCbar \Pr(\calCbar_i) + \gDC \Pr(\calC_i) \nonumber \\
	& ~~~~~~~~~~~~~ = \begin{cases}
		j\alpha_i, & \text{if $0 \le j < 1/\alpha_i$}\\
		1 - \alpha_i \frac{n}{M}, & \text{if $j \ge 1/\alpha_i$}
	\end{cases}. \label{eq:partg}
\end{align}

\section{Minimizing the Weighted Communication Rate}\label{sec:min}
We consider the optimization of the content allocation $\ba$. Similar to the work in \cite{Shanmugam2013}, where the average delay, formulated as a linear scalarization of the macro BS and small BS download delays, we minimize the \emph{weighted communication rate}
\begin{equation}\label{eq:h}
	h(\ba) \triangleq \theta f(\ba) + (1-\theta) g(\ba)
\end{equation}
for some given $\theta$, $0.5 \le \theta \le 1$. Note that the communication rate is directly related to the download delay considered in \cite{Shanmugam2013}. Minimizing the expected downlink rate corresponds to $\theta = 1$. However, it might be desirable to also limit the D2D communication rate for various reasons, such as device battery constraints and interference between devices. Therefore, we consider $0.5 \le \theta \le 1$, where $\theta \ge 0.5$ stems from the fact that the bottleneck is in the downlink. The weighted communication rate \eqref{eq:h} can be rewritten as shown in \eqref{eq:hexpand} at the top of the page using \eqref{eq:f}--\eqref{eq:partg}. The minimization of the weighted communication rate \eqref{eq:h} in terms of the content allocation can then be formulated as the following optimization problem
\addtocounter{equation}{1}
\begin{subequations}
\label{P1}
	\begin{align}
		\underset{\alpha_i \in \calA}{\text{minimize}}~ & h(\ba) \label{obj:h}\\
		\text{subject to}~ & \sum_{i = 1}^N \alpha_i \le \beta. \label{cnstr:beta}
	\end{align}
\end{subequations}
We denote by $\ba^*$ the \emph{optimal} content allocation resulting from \eqref{P1}. In the following theorem, we rewrite the optimization problem \eqref{P1} in an equivalent form that is tractable.

\begin{theorem}
Problem (\ref{P1}a)--(\ref{P1}b) is equivalent to the MILP
\begin{subequations}\label{P2}
	\begin{align}
		\underset{\substack{\alpha_i \in \mathbb{R}\\t_{ij} \in \mathbb{R}\\ b_{i\ell} \in \{0, 1\}}}{\textnormal{minimize}}~ & \sum_{i=1}^N \sum_{j=0}^\infty t_{ij}\label{obj:t}\\
		\textnormal{subject to}~ & \sum_{i = 1}^N \alpha_i \le \beta\\
			& t_{ij} + p_i q_j \left((2\theta-1)j+\theta\frac{n}{M}\right)\alpha_i \ge \theta p_i q_j\label{cnstr:epi1}\\
			& t_{ij} + (1-\theta) \frac{n}{M} p_i q_j \alpha_i \ge (1-\theta) p_i q_j\label{cnstr:epi2}\\
			& \alpha_i - \sum_{\ell=1}^n \frac{b_{i\ell}}{n-\ell+1} = 0\label{cnstr:b1}\\
			& \sum_{\ell=0}^n b_{i\ell} = 1\label{cnstr:b2}
	\end{align}
\end{subequations}
\end{theorem}
\begin{IEEEproof}
The objective function \eqref{eq:hexpand} is a sum of piecewise linear functions of $\alpha_i$. This allows us to rewrite the optimization problem in a way that is tractable, using the epigraph formulation \cite{Boyd2009}. Using \eqref{eq:hexpand} and introducing a new optimization variable $t_{ij} \in \mathbb{R}$, we minimize the objective function \eqref{obj:t} over the optimization variables $\alpha_i\in\calA$ and $t_{ij}\in\mathbb{R}$. The constraints \eqref{cnstr:epi1} and \eqref{cnstr:epi2}, which arise from the first and second term in the $\max$ function in \eqref{eq:hexpand}, are added to the optimization problem. Note that we drop the factor $M\omega$ in \eqref{obj:t} since it is irrelevant to the solution of the optimization problem. Variables $\{\alpha_i\}$ can only take on the discrete values given by \eqref{eq:alpha}. To handle this, we introduce the binary auxiliary optimization variable $b_{i\ell} \in \{0, 1\}$ and let
\begin{equation}
\label{eq:newalpha}
	\alpha_i = b_{i0} \cdot 0 + b_{i1}\frac{1}{n} + b_{i2}\frac{1}{n-1} + \ldots + b_{in},~\forall i,
\end{equation}
which constitutes constraint \eqref{cnstr:b1} \cite[Sec.~3.2]{Appa2006}. To guarantee that, for each $i$, only one $b_{i\ell}\neq0$, the constraint \eqref{cnstr:b2} is added to the problem. We can now optimize over the variable $\alpha_i\in\mathbb{R}$ and formulate the MILP in \eqref{P2} that is equivalent to \eqref{P1}.
\end{IEEEproof}

So far, an average cache size constraint has been assumed in order to simplify the optimization problems \eqref{P1} and \eqref{P2}. We suggest the following greedy approach to enforce a strict cache size constraint per device. For a cache size overhead, denoted by $\delta \ge 0$, Algorithm~\ref{alg:strict} enforces strictly the cache size constraint $(1+\delta)\bbar$ per device. We refer to the output of the algorithm as the \emph{strict} content allocation and denote it by $\ba'$.

For $N\le100$ and $n=M\le1000$ (approximately), we are able to solve the optimization problem in \eqref{P2} using a branch-and-bound method with a guarantee to attain the best bound, i.e., the optimality gap goes to zero. To be able to solve for a larger code length $n$, i.e., potentially a larger number of devices $M$, we consider a relaxation of the optimization problem in \eqref{P2} where the integer constraints \eqref{cnstr:b1} and \eqref{cnstr:b2} of the MILP \eqref{P2} are replaced by the constraint $0 \le \alpha_i \le 1$, resulting in the LP
\begin{align}
	\underset{\substack{\alpha_i \in \mathbb{R}\\t_{ij} \in \mathbb{R}}}{\textnormal{minimize}}~ & \sum_{i=1}^N \sum_{j=0}^\infty t_{ij}\label{P3}\\
	\textnormal{subject to}~ & \sum_{i = 1}^N \alpha_i \le \beta \nonumber\\
		& t_{ij} + p_i q_j \left((2\theta-1)j+\theta\frac{n}{M}\right)\alpha_i \ge \theta p_i q_j \nonumber\\
		& t_{ij} + (1-\theta) \frac{n}{M} p_i q_j \alpha_i \ge (1-\theta) p_i q_j \nonumber\\
		& 0 \le \alpha_i \le 1\nonumber.
\end{align}
We denote by $\arlx$ the allocation resulting from \eqref{P3} and refer to it as the \emph{integer-relaxed optimal} content allocation. Note that the weighted communication rate \eqref{eq:h} resulting from the integer-relaxed optimal allocation is at least as good as the weighted communication rate using the optimal content allocation provided by the MILP solution \eqref{P2} since the allocation obtained from \eqref{P2} is also a feasible solution to \eqref{P3}. Hence, the weighted communication rate using the allocation obtained from \eqref{P3} is a lower bound on the weighted communication rate using the allocation obtained from \eqref{P2}, i.e.,
\begin{equation}\label{eq:lb}
	h(\arlx) \le h(\ba^*).
\end{equation}
In Section~\ref{sec:results}, we observe numerically that $h(\ba^*)\approx h(\arlx)$, i.e., the weighted communication rate resulting from the integer-relaxed content allocation $\arlx$ represents well the weighted communication rate corresponding to the optimal content allocation $\ba^*$.

\begin{algorithm}[t!]
\caption{Greedy strict cache size constraint}\label{alg:strict}
\begin{algorithmic}[1]
	\REQUIRE $\ba = \left(\alpha_1, \ldots, \alpha_N\right)$, $N$, $M$, $n$, $\bbar$, and $\delta$
	\ENSURE $\astr = \left(\alpha'_1, \ldots, \alpha'_N\right)$
	\STATE $\astr \leftarrow \ba$
	\STATE $c_{ij} \leftarrow 0,~i = 1, \ldots, N,~j = 1, \ldots, M$
	\STATE $i \leftarrow 1$
	\WHILE{$i \le N$}
		\STATE $\mathcal{M} \leftarrow \{1, \ldots, M\}$, $\ell \leftarrow n$
		\WHILE{$\ell>0$}
			\STATE $\mathcal{L} \leftarrow \ell$ values of $\mathcal{M}$, chosen uniformly at random without replacement
			\FOR{$j \in \{j:j \in \mathcal{L}, \alpha'_i + \sum_{m=1}^{i-1} \alpha'_m c_{mj} \le (1+\delta)\bbar\}$}
				\STATE $c_{ij} \leftarrow 1$
			\ENDFOR
			\STATE $\mathcal{M} \leftarrow \mathcal{M}\backslash \mathcal{L}$
			\STATE $\ell \leftarrow n-\sum_{j=1}^M c_{ij}$
			\IF{$\ell > |\mathcal{M}|$}
				\STATE \textbf{break}
			\ENDIF
		\ENDWHILE
		\IF{$\sum_{j=1}^M c_{ij} < n$}
			\STATE $c_{ij} \leftarrow 0,~j = 1, \ldots, M$
			\STATE $k_i \leftarrow \frac{1}{\alpha'_i}+1$
			\IF{$k_i > n$}
				\STATE $\alpha'_i \leftarrow 0$
			\ELSE
				\STATE $\alpha'_i \leftarrow 1/k_i$
			\ENDIF
		\ELSE
			\STATE $i \leftarrow i+1$
		\ENDIF
	\ENDWHILE
\end{algorithmic}
\end{algorithm}

\begin{algorithm}[t!]
\caption{Round-to-integer content allocation}\label{alg:round}
\begin{algorithmic}
	\REQUIRE $\arlx = \left(\check{\alpha}_1, \ldots, \check{\alpha}_N\right)$, $n$, and $N$
	\ENSURE $\afloor = \left(\hat{\alpha}_1, \ldots, \hat{\alpha}_N\right)$
	\FOR{$i=1,\ldots,N$}
		\STATE $\hat{\alpha}_i \leftarrow \frac{1}{\left\lceil 1/\check{\alpha}_i \right\rceil}$
		\IF{$\hat{\alpha}_i<1/n$}
			\STATE $\hat{\alpha}_i \leftarrow 0$
		\ENDIF
	\ENDFOR
\end{algorithmic}
\end{algorithm}
Note that a practical coding scheme must have valid values of $\alpha_i$, i.e., $\alpha_i\in\calA$. In other words, valid values of $\alpha_i$ are such that $k_i=1/\alpha_i$ is integer in $[1,n]$ and therefore a code $(n,k_i)$ can be realized. This is not guaranteed by the solution of the LP \eqref{P3}. To remedy this problem, we suggest a simple algorithm to ensure valid values of $\alpha_i$ from the values $\checka$ resulting from \eqref{P3}, without violating the cache size constraint \eqref{eq:beta}. The algorithm is given in Algorithm~\ref{alg:round}. We denote the resulting content allocation by $\afloor$ and refer to it as the \emph{round-to-integer} content allocation. By using Algorithm~\ref{alg:round}, we are guaranteed valid values of $\alpha_i$. Note that the weighted communication rate arising from the content allocation provided by Algorithm~\ref{alg:round} is higher than or equal to the weigthed communication rate with the optimal content allocation $\ba^*$ obtained solving \eqref{P2}. Therefore, the weighted communication rate using the allocation provided by Algorithm~\ref{alg:round} is an upper bound to the weighted communication rate using the optimal content allocation, i.e.,
$$
	h(\afloor) \ge h(\ba^*).
$$
Using \eqref{eq:lb}, we have
\begin{equation}\label{eq:sandwitch}
	h(\arlx) \le h(\ba^*) \le h(\afloor).
\end{equation}

As shown in Section~\ref{sec:results}, our numerical results show that, for the important case of maximal spreading, the gap between the upper and lower bounds is very small.

For the specific case of equally expensive downlink and D2D communication, i.e., $\theta = 0.5$, it is optimal to use the popular content allocation \eqref{eq:apop}. The result is given by the following proposition.

\begin{proposition}\label{prop:popopt}
	For $\theta = 0.5$ and $\beta \in \mathbb{N}$, popular content allocation \eqref{eq:apop} is optimal.
\end{proposition}
\begin{IEEEproof}
	For $\theta = 0.5$, \eqref{eq:hexpand} reduces to
	\begin{align*}
		h(\ba) & = \frac{M\omega}{2} \sum_{i=1}^N p_i \sum_{j=0}^\infty q_j \left(1-\alpha_i\frac{n}{M}\right)\\
			& = \frac{M\omega}{2} \left(1 - \frac{n}{M} \sum_{i=1}^N \alpha_i p_i\right)
	\end{align*}
	and the minimization problem \eqref{P1} converts to the maximization problem
	\begin{subequations}
		\begin{align}
			\underset{\alpha_i \in \calA}{\text{maximize}}~ & \sum_{i=1}^N \alpha_i p_i \label{obj:max}\\
			\text{subject to}~ & \sum_{i = 1}^N \alpha_i \le \beta \label{cnstr:max}.
		\end{align}
	\end{subequations}
	Since $p_1 \ge p_2 \ge \ldots \ge p_N$ according to \eqref{eq:zipf}, it is trivial to see that the sum \eqref{obj:max} is maximized for $\alpha_i = 1,~i=1, \ldots, \beta$, i.e., it is optimal to use the popular content allocation.
\end{IEEEproof}

\section{Numerical results}\label{sec:results}
\begin{figure}[t!]
	\centering
	\begin{tikzpicture}[>=stealth,
	]
\begin{semilogxaxis}[
	width = \columnwidth,
%	scaled ticks = false,
%	tick label style = {/pgf/number format/fixed},
	xmin=0.01,
	xmax=1,
	xlabel = {$n/M$},
	ymin=25,
	ymax=50,
	ylabel = {$f(\ba)$},
	legend cell align = left,
	legend pos = south west,
]

\addplot [color = black, densely dashdotted, forget plot]
table[row sep=crcr]{%
		0.0100	43.0930\\
		0.0200	40.1533\\
		0.0300	38.5295\\
		0.0400	36.8962\\
		0.0500	35.9018\\
		0.0600	35.1803\\
		0.0700	34.6511\\
		0.0800	34.2104\\
		0.0900	33.8518\\
		0.1000	33.9331\\
		0.2000	31.7007\\
		0.3000	30.9788\\
		0.4000	30.6460\\
		0.5000	30.3952\\
		0.6000	30.2474\\
		0.7000	30.1546\\
		0.8000	30.1217\\
		0.9000	30.1812\\
		1.0000	29.0453\\
};
\addplot [color = color3, solid, forget plot]
	table[row sep=crcr]{%
		0.002	48.5357\\
		0.004	47.1140\\
		0.008	44.3939\\
		0.01		43.0930\\
		0.02		40.1702\\
		0.04		37.1358\\
		0.05		36.3393\\
		0.1		35.3619\\
		0.2		37.4286\\
		0.25		38.5826\\
		0.5		42.3191\\
		1		45.2434\\
	};
\addplot [color = color2, solid, forget plot]
	table[row sep=crcr]{%
		0.0100	46.7910\\
		0.0200	44.2311\\
		0.0300	42.6603\\
		0.0400	41.3677\\
		0.0500	40.2544\\
		0.0600	39.3448\\
		0.0700	38.5550\\
		0.0800	37.8444\\
		0.0900	37.2257\\
		0.1000	36.6997\\
		0.2000	33.5320\\
		0.3000	31.9823\\
		0.4000	31.0879\\
		0.5000	30.4626\\
		0.6000	30.0256\\
		0.7000	29.6887\\
		0.8000	29.4327\\
		0.9000	29.2302\\
		1.0000	29.0787\\
	};
\addplot [color = color1, densely dashed, forget plot]
table[row sep=crcr]{%
		0.0100	42.6270\\
		0.0200	40.0599\\
		0.0300	38.0984\\
		0.0400	36.7127\\
		0.0500	35.6928\\
		0.0600	34.9195\\
		0.0700	34.3202\\
		0.0800	33.8465\\
		0.0900	33.4603\\
		0.1000	33.1437\\
		0.2000	31.4129\\
		0.3000	30.6203\\
		0.4000	30.1503\\
		0.5000	29.8264\\
		0.6000	29.5916\\
		0.7000	29.4096\\
		0.8000	29.2691\\
		0.9000	29.1496\\
		1.0000	29.0442\\
};
\addplot [color = color1, solid, forget plot]
	table[row sep=crcr]{%
		0.01	43.093\\
		0.02	40.1135\\
		0.05	35.8494\\
		0.1	33.3751\\
		0.2	31.5522\\
		0.5	29.8879\\
		1	29.0453\\
	};
\draw[densely dotted] (axis cs:0.08, 35.3619) to (axis cs:0.9, 35.3619);
\draw[densely dotted] (axis cs:0.7, 29.0453) to (axis cs:1, 29.0453);
\draw[<-] (axis cs:0.8, 29.0453) to node[left, near end]{$\sim18\%$} (axis cs:0.8, 35.3619);
\addplot [color = color2, only marks, every mark/.append style = {fill = white, scale = 0.9}, mark = *, forget plot]
	table[row sep=crcr]{%
		0.0100	46.9030\\
		0.0200	44.275\\
		0.0500	40.2931\\
		0.1000	36.7266\\
		0.2000	33.5808\\
		0.5000	30.5092\\
	};
\addplot [color = color3, only marks, every mark/.append style = {fill = white, scale = 1.2}, mark = triangle*, forget plot]
	table[row sep=crcr]{%
		0.05	36.2735\\
		0.1	35.4136\\
		0.2	37.5239\\
%		0.25	38.6053\\
		0.5	42.1972\\
		1	45.24341\\
	};
\addplot [color=color1, only marks, every mark/.append style = {fill = white, scale = .8}, mark = square*, forget plot]
	table[row sep=crcr]{%
		0.01	43.1925\\
		0.02	40.0525\\
		0.05	35.8255\\
		0.1	33.2913\\
		0.2	31.3656\\
		0.5	29.7823\\
		1	29.0989\\
	};
\addlegendimage{color = color1, every mark/.append style = {fill = white, scale = .8}, mark = square*};
\addlegendentry{$\ba^*$};
\addlegendimage{color = color1, densely dashed}
\addlegendentry{$\arlx$}
\addlegendimage{color = color2, every mark/.append style = {fill = white, scale = 0.9}, mark = *};
\addlegendentry{$\afloor$};
\addlegendimage{color = color3, every mark/.append style = {fill = white, scale = 1.2}, mark = triangle*};
\addlegendentry{$\apop$};
\addlegendimage{color = black, densely dashdotted};
\addlegendentry{$\astr$};
\end{semilogxaxis}%
\end{tikzpicture}%
	\vspace{-3ex}
	\caption{The downlink rate $f(\ba)$ versus $n/M$ using different content allocations for $M = 500$, $\sigma = 0.7$, $\theta = 1$, and $\bbar = 1$. All markers correspond to simulated downlink rate.}
	\label{fig:spreadlow}
	\vspace{-3ex}
\end{figure}
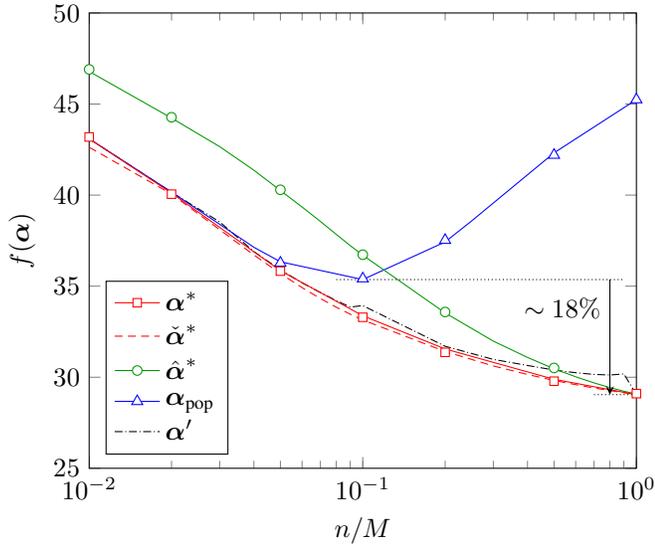

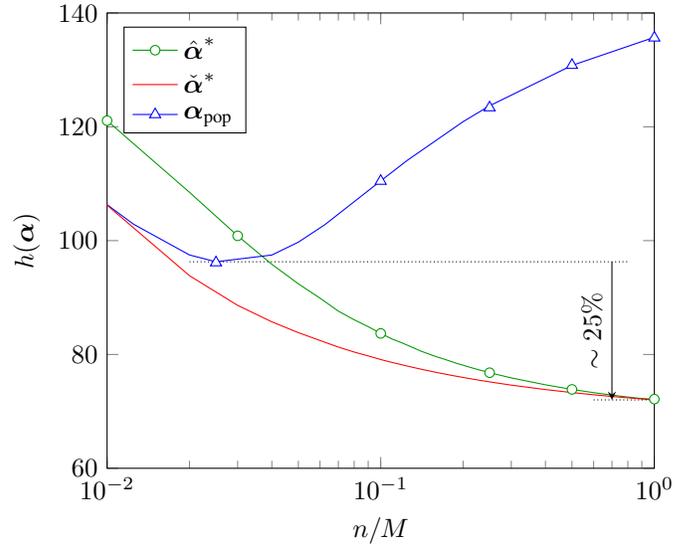
\begin{figure}[t!]
	\centering
	\begin{tikzpicture}[>=stealth]
\begin{semilogxaxis}[%
	width = \columnwidth,
	scaled ticks = false,
	tick label style = {/pgf/number format/fixed},
%	log ticks with fixed point,
	xmin=0.01,
	xmax=1,
	xlabel = {$n/M$},
	ymin=60,
	ymax=140,
	ylabel = {$h(\ba)$},
	legend cell align = left,
	legend pos = north west,
]

\addplot [color = color2, solid, forget plot]
	table[row sep=crcr]{%
		0.0100	121.0773\\
		0.0200	108.5015\\
		0.0300	100.8471\\
		0.0400	 95.8449\\
		0.0500	 92.4127\\
		0.0600	 89.8990\\
		0.0700	 87.6169\\
		0.0800	 86.0911\\
		0.0900	 84.8878\\
		0.1000	 83.6967\\
		0.1100	 82.7179\\
		0.1200	 82.0065\\
		0.1300	 81.2871\\
		0.1400	 80.6037\\
		0.1500	 80.0751\\
		0.1600	 79.5914\\
		0.1700	 79.2035\\
		0.1800	 78.8027\\
		0.1900	 78.4144\\
		0.2000	 78.1169\\
		0.2100	 77.7979\\
		0.2200	 77.5215\\
		0.2300	 77.2538\\
		0.2400	 77.0413\\
		0.2500	 76.7917\\
		0.3000	 75.8843\\
		0.4000	 74.6692\\
		0.5000	 73.8575\\
		0.6000	 73.2931\\
		0.7000	 72.8580\\
		0.7500	 72.6805\\
		0.8000	 72.5213\\
		0.9000	 72.2473\\
		1.0000	 72.1498\\
	};

\addplot [color = color3, solid, forget plot]
	table[row sep=crcr]{%
		0		200\\
		0.0005	147.1882\\
		0.0010	144.4540\\
		0.0020	139.2099\\
		0.0025	136.6959\\
		0.0040	129.5623\\
		0.0050	125.1282\\
		0.0080	113.2192\\
		0.0100	106.3174\\
		0.0125	102.8799\\
		0.0200	97.4860\\
		0.0250	96.2873\\
		0.0400	97.4715\\
		0.0500	99.7321\\
		0.0625	102.8029\\
		0.1000	110.5000\\
		0.1250	114.1000\\
		0.2000	120.9000\\
		0.2500	123.6941\\
		0.5000	130.7885\\
		1.0000	135.7302\\
	};

\addplot [color = color1, solid, forget plot]
	table[row sep=crcr]{%
		0		200\\
		0.0005	133.8790\\
		0.0010	132.0724\\
		0.0020	128.5893\\
		0.0025	126.9111\\
		0.0040	122.1197\\
		0.0050	119.1199\\
		0.0080	110.9860\\
		0.0100	106.2268\\
		0.0200	93.8451\\
		0.0300	88.6294\\
		0.0400	85.7351\\
		0.0500	83.8373\\
		0.0600	82.4514\\
		0.0700	81.3005\\
		0.0800	80.4230\\
		0.0900	79.7426\\
		0.1000	79.1105\\
		0.1100	78.5965\\
		0.1200	78.1682\\
		0.1300	77.7678\\
		0.1400	77.4301\\
		0.1500	77.1203\\
		0.1600	76.8371\\
		0.1700	76.5954\\
		0.1800	76.3595\\
		0.1900	76.1554\\
		0.2000	75.9611\\
		0.2100	75.7808\\
		0.2200	75.6214\\
		0.2300	75.4613\\
		0.2400	75.3234\\
		0.2500	75.1817\\
		0.3000	74.6153\\
		0.4000	73.8333\\
		0.5000	73.3059\\
		0.6000	72.9187\\
		0.7000	72.6210\\
		0.75		72.4947
		0.8000	72.3823\\
		0.9000	72.1852\\
		1.0000	72.0189\\
};
\draw[densely dotted] (axis cs:0.02, 96.2873) to (axis cs:0.8, 96.2873);
\draw[densely dotted] (axis cs:0.6, 72.0189) to (axis cs:1, 72.0189);
\draw[<-] (axis cs:0.7, 72.0189) to node[above, midway, sloped]{$\sim25\%$} (axis cs:0.7, 96.2873);

\addplot [color = color2, only marks, every mark/.append style = {fill = white, scale = 0.9}, mark = *, forget plot]
	table[row sep=crcr]{%
		0.0100	121.0773\\
		0.0300	100.8471\\
		0.1000	 83.6967\\
		0.2500	 76.7917\\
		0.5000	 73.8575\\
%		0.7500	 72.6805\\
		1.0000	 72.1498\\
	};
\addplot [color = color3, only marks, every mark/.append style = {fill = white, scale = 1.2}, mark = triangle*, forget plot]
	table[row sep=crcr]{%
		0.025	96.137\\
		0.1		110.460\\
		0.25		123.373\\
		0.5		130.827\\
		1		135.635\\
	};
\addlegendimage{color = color2, every mark/.append style = {fill = white, scale = 0.9}, mark = *};
\addlegendentry{$\afloor$};
\addlegendimage{color = color1};
\addlegendentry{$\arlx$};
\addlegendimage{color = color3, every mark/.append style = {fill = white, scale = 1.2}, mark = triangle*};
\addlegendentry{$\apop$};
\end{semilogxaxis}%
\end{tikzpicture}%
	\vspace{-3ex}
	\caption{The weighted communication rate $h(\ba)$ versus $n/M$ using different content allocations for $M = 2000$, $\sigma = 0.7$, $\theta = 0.75$, and $\bbar = 1$. All markers correspond to simulated rate.}
	\label{fig:spreadhigh}
	\vspace{-3ex}
\end{figure}

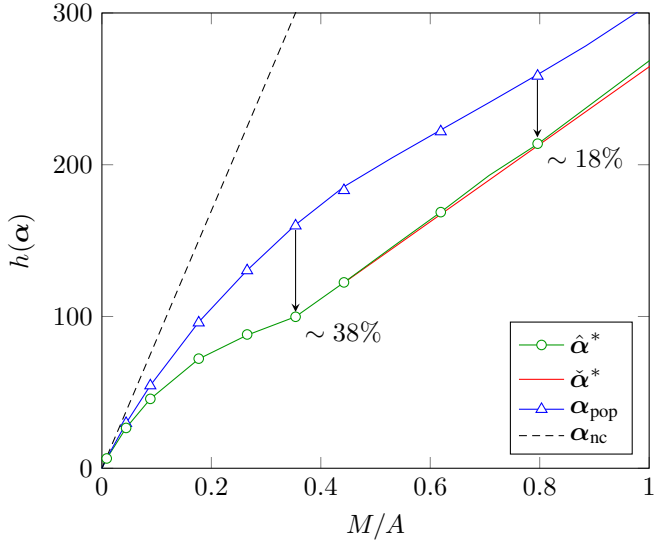
\begin{figure}[t!]
	\centering
	\begin{tikzpicture}[>=stealth]
\begin{axis}[%
	width = \columnwidth,
	xmin = 0,
	xmax = 1,
	xlabel = {$M/A$},
	ymin = 0,
	ymax = 300,
%	axis y line* = left,
	ylabel = {$h(\ba)$},
	legend cell align = left,
	legend pos = south east,
]
\addplot [color = color1, solid, forget plot]
	table[row sep=crcr]{%
		0.4423  122.2382\\
		0.5308  144.7814\\
		0.6192  167.3366\\
		0.7077  189.9013\\
		0.7962  212.4733\\
		0.8846  235.0513\\
		1.3270  347.9996\\
		1.7693  461.0037\\
};
% Floor
\addplot[color = color2, solid, forget plot]
	table[row sep = crcr]{%
		0	0\\
		0.0088	6.3868\\
		0.0442	26.6157\\
		0.0885	45.6778\\
		0.1769	72.0179\\
		0.2654	87.7324\\
		0.3539	99.8646\\
		0.4423  122.4168\\
		0.5308  145.8080\\
		0.6192  168.8065\\
		0.7077  192.9944\\
		0.7962  213.6080\\
		0.8846  237.4064\\
		1.3270  356.4139\\
		1.7693  475.4514\\
	};
\addplot [color = color3, solid, forget plot]
	table[row sep=crcr]{%
		0	0\\
		0.0088	6.6256\\
		0.0442	29.7087\\
		0.0884	54.4998\\
		0.1768	96.2873\\
		0.2653  130.4581\\
		0.3537  160.3896\\
		0.4421  185.4818\\
		0.5305  204.6116\\
		0.6189  222.8364\\
		0.7074  240.9257\\
		0.7958  259.3290\\
		0.8842  278.2859\\
		1.3263  382.7648\\
		1.7684  498.8000\\
	};

% No caching
\addplot [color = black, densely dashed, forget plot]
	table[row sep=crcr]{%
		0	0\\
		0.0088	7.5\\
		0.0442	37.5\\
		0.0884	75\\
		0.1768	150\\
		0.2653	225\\
		0.3537	300\\
		0.4421	375\\
		0.5305	450\\
		0.6189	525\\
		0.7074	600\\
		0.7958	675\\
		0.8842	750\\
		1.3263	1125\\
		1.7684	1500\\
	};
\draw[<-] (axis cs:0.3539, 103) to node[below, at start, anchor = north west]{$\sim38\%$} (axis cs:0.3539, 157);
\draw[<-] (axis cs:0.7958, 218) to node[below, at start, anchor = north west]{$\sim18\%$} (axis cs:0.7958, 256);
%
% Simulations >>>>>>>>>>>>>>>>>>>>>>>
\addplot [color = color3, only marks, every mark/.append style = {fill = white, scale = 1.2}, mark = triangle*, forget plot]
	table[row sep=crcr]{%
%		0.0088	6.5316\\ % M = 100
		0.0442	29.6324\\ % M = 500
  		0.0885	54.4443\\ % M = 1000
		0.1769	95.885\\ % M = 2000
		0.2654	130.2562\\ % M = 3000
		0.3539	159.7959\\ % M = 4000
		0.4423	183.0845\\ % M = 5000
%		0.5308	\\ % M = 6000
		0.6192	221.8175\\ % M = 7000
		0.7962	258.4621\\
		1.3270	382.43191\\ % M = 15 000
	};
\addplot [color=color2, only marks, every mark/.append style = {fill = white, scale = 0.9}, mark = *, forget plot]
	table[row sep=crcr]{%
		0.0088	6.4566\\ % M = 100
		0.0442	26.5675\\ % M = 500
  		0.0885	45.72\\ % M = 1000
		0.1769	72.23162\\ % M = 2000
		0.2654	88.15775\\ % M = 3000
		0.3539	99.8081\\ % M = 4000
		0.4423	122.4667\\ % M = 5000
		0.6192	168.7587\\ % M = 7000
		0.7962	213.8732\\ % M = 9000
	};
	\addlegendimage{color = color2, every mark/.append style = {fill = white, scale = 0.9}, mark = *};
	\addlegendentry{$\afloor$};
	\addlegendimage{color = color1};
	\addlegendentry{$\arlx$};
	\addlegendimage{color = color3, every mark/.append style = {fill = white, scale = 1.2}, mark = triangle*};
	\addlegendentry{$\apop$};
	\addlegendimage{color = black, densely dashed};
	\addlegendentry{$\anc$};
\end{axis}
\end{tikzpicture}%
	\vspace{-3ex}
	\caption{The weighted communication rate $h(\ba)$ versus the device density $M/A$ using different content allocations for $\sigma = 0.7$, $\theta = 0.75$, and $\bbar = 1$. All markers correspond to simulated rate.}
	\label{fig:density}
	\vspace{-3ex}
\end{figure}

\begin{figure}[t!]
	\centering
	\begin{tikzpicture}[>=stealth]
\begin{axis}[%
	width = \columnwidth,
	xmin = 0.5,
	xmax = 1,
	xlabel = {$\theta$},
	ymin = 0,
	ymax = 500,
%	axis y line* = left,
	ylabel = {$h(\ba)$},
	legend cell align = left,
	legend pos = north west,
]
%
% M = 2000
\addplot [color = color1, solid, forget plot]
	table[row sep=crcr]{%
		0.5	90.4868\\
		0.5100	92.0128\\
		0.5200	92.9070\\
		0.5300	93.3078\\
		0.5400	93.2848\\
		0.5500	92.8750\\
		0.5600	92.1024\\
		0.5700	91.0657\\
};
\addplot [color = color2, solid, forget plot]
	table[row sep=crcr]{%
		0.5000	90.4868\\
		0.5100	96.0532\\
		0.5200	95.7948\\
		0.5300	94.9163\\
		0.5400	93.3820\\
		0.5500	93.4944\\
		0.5600	92.2092\\
		0.5700	91.1161\\
		0.5800	90.1009\\
		0.5900	89.0161\\
		0.6000	87.9635\\
		0.7000	77.4108\\
		0.7500	72.1498\\
		0.8000	66.8023\\
		0.9000	56.3147\\
		1.0000	45.6840\\
	};
\addplot [color = color3, solid, forget plot]
	table[row sep=crcr]{%
		0.5000	90.4868\\
		0.5100	92.2966\\
		0.5200	94.1063\\
		0.5300	95.9160\\
		0.5400	97.6683\\
		0.5500	97.6589\\
		0.5600	97.6496\\
		0.5700	97.6402\\
		0.5800	97.6249\\
		0.5900	97.5462\\
		0.6000	97.4675\\
		0.7000	96.6807\\
		0.7500	96.2873\\
		0.8000	95.8939\\
		0.9000	95.1071\\
		1.0000	94.3204\\
	};
\addplot [color = black, densely dashed, forget plot]
	table[row sep=crcr]{%
		0.5	100\\
		0.6	120\\
		0.7	140\\
		0.75	150\\
		0.8	160\\
		0.9	180\\
		1	200\\
	};
%
% M = 5000
%
\addplot [color = color1, solid, forget plot]
	table[row sep=crcr]{%
		0.5	226.217\\
		0.5100	228.9079\\
		0.5200	228.7996\\
		0.5300	226.7075\\
		0.5400	222.9451\\
		0.5500	218.3620\\
		0.5600	213.6994\\
		0.5700	208.9921\\
		0.5800	204.2550\\
		0.5900	199.4964\\
		0.6	194.7213\\
	};
\addplot [color = color2, solid, forget plot]
	table[row sep=crcr]{%
		0.5000	226.2170\\
		0.5100	238.3031\\
		0.5200	234.4954\\
		0.5300	227.7894\\
		0.5400	224.2708\\
		0.5500	219.1547\\
		0.5600	214.1699\\
		0.5700	209.2433\\
		0.5800	204.3192\\
		0.5900	200.4147\\
		0.6000	195.4842\\
		0.7000	146.9566\\
		0.7500	122.4670\\
		0.8000	97.9189\\
		0.9000	49.1409\\
		1.0000	0.0021\\
	};
\addplot [color = color3, solid, forget plot]
	table[row sep=crcr]{%
		0.5000	226.2170\\
		0.5100	230.7414\\
		0.5200	235.2657\\
		0.5300	239.7901\\
		0.5400	237.5771\\
		0.5500	235.0964\\
		0.5600	232.6156\\
		0.5700	230.1349\\
		0.5800	227.6542\\
		0.5900	225.1735\\
		0.6000	222.6927\\
		0.7000	197.8855\\
		0.7500	185.4818\\
		0.8000	173.0782\\
		0.9000	148.2709\\
		1.0000	123.4636\\
	};
\addplot [color = black, densely dashed, forget plot]
	table[row sep=crcr]{%
		0.5	250\\
		0.6	300\\
		0.7	350\\
		0.75	375\\
		0.8	400\\
		0.9	450\\
		1	500\\
	};

% M = 2000
\addplot [color = color3, only marks, every mark/.append style = {fill = white, scale = 1.2}, mark = triangle*, forget plot]
	table[row sep=crcr]{%
		0.6	97.5143\\
		0.75	96.2374\\
		0.9	95.1071\\
		1	94.3204\\
	};
\addplot [color=color2, only marks, every mark/.append style = {fill = white, scale = .9}, mark = *, forget plot]
	table[row sep=crcr]{%
		0.5	90.3304\\
		0.6	88.07418\\
		0.75	72.23162\\
		0.9	56.53745\\
		1	46.2126\\
	};
%
% M = 5000
%
\addplot [color = color3, only marks, every mark/.append style = {fill = white, scale = 1.2}, mark = triangle*, forget plot]
	table[row sep=crcr]{%
		0.6	222.6927\\
		0.75	185.4818\\
		0.9	148.2709\\
		1	123.4636\\
	};
\addplot [color=color2, only marks, every mark/.append style = {fill = white, scale = .9}, mark = *, forget plot]
	table[row sep=crcr]{%
		0.5	226.4288\\
		0.6	194.7919\\
		0.75	122.4667\\
		0.9	49.0362\\
		1	0.0013\\
	};
\draw (axis cs:0.6, 250) ellipse (10 and 80);
\node at (axis cs:0.68, 260) {$M=5000$};
\draw (axis cs:0.6, 100) ellipse (10 and 40);
\node at (axis cs:0.6, 40) {$M=2000$};
\addlegendimage{color = color2, every mark/.append style = {fill = white, scale = 0.9}, mark = *};
\addlegendentry{$\afloor$};
\addlegendimage{color = color1};
\addlegendentry{$\arlx$};
\addlegendimage{color = color3, every mark/.append style = {fill = white, scale = 1.2}, mark = triangle*};
\addlegendentry{$\apop$};
\addlegendimage{color = black, densely dashed};
\addlegendentry{$\anc$};
%\addlegendimage{color = black};
%\addlegendentry{Gain};
\end{axis}
\end{tikzpicture}%
	\vspace{-3ex}
	\caption{The weighted communication rate $h(\ba)$ versus the weight $\theta$ for $\sigma = 0.7$ and $\bbar = 1$. All markers correspond to simulated rate.}
	\label{fig:weight}
%	\vspace{-1ex}
\end{figure}
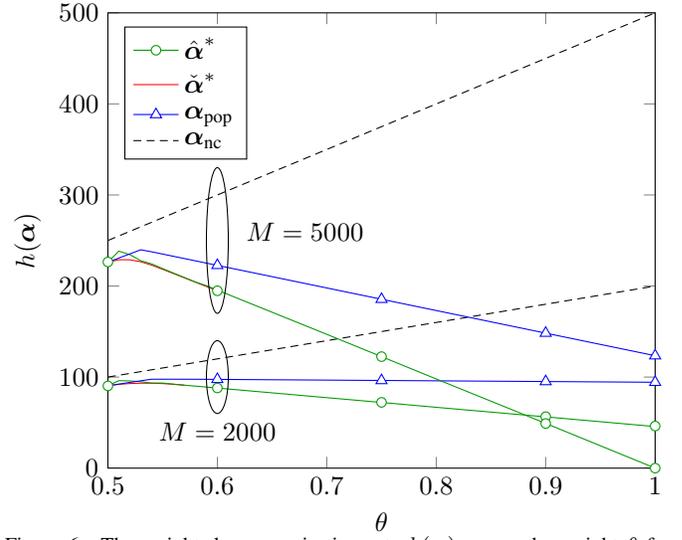

In Figs.~\ref{fig:spreadlow}--\ref{fig:sigma}, we evaluate the downlink rate \eqref{eq:f} and the weighted communication rate \eqref{eq:h} for the optimal content allocation $\ba^*$ obtained by solving the MILP \eqref{P2}, with lower and upper bounds given by the integer-relaxed optimal content allocation $\arlx$ provided by the solution of the LP \eqref{P3}, and the round-to-integer content allocation $\afloor$ provided by Algorithm~\ref{alg:round}, respectively. Specifically, we investigate the reduction in the weighted communication rate of using the optimal content allocation over the popular content allocation, for which we consider a code length $n$ such that $\beta \in \mathbb{N}$, i.e., the constraint \eqref{eq:beta} is attained with equality, which means that the content allocation uses all the available cache space. For the results, we consider a file library with $N = 100$ files, an area on a sphere of radius $\rho = 30$ meters, which corresponds to an area of roughly $11000$ square meters, and a communication range $r = 10$ meters. These values of $\rho$ and $r$ are enough to satisfy the condition $r\ll2\rho$ and provide a good approximation of the distribution of devices within the communication range in Theorem~\ref{thm:qj}. This is verified by computing the Kullback-Leibler divergence between the empirical distribution of the number of caching devices within the communication range of any reference device, obtained through simulations, and the theoretical distribution provided by Theorem~\ref{thm:qj}. To reflect a typical walking speed, we assume a minimum and a maximum device speed of $\smin = 0.3$ and $\smax = 2.5$ m/s, respectively, and a request rate $\omega = 0.1$ $\text{s}^{-1}$. Note that the optimal content allocation does not depend on $\omega$. In Figs.~\ref{fig:spreadlow}--\ref{fig:bbar}, we set $\sigma=0.7$ motivated by the frequency of document accesses \cite{Breslau1999} and the popularity of YouTube videos under the assumption that the video popularity follows the Zipf distribution  \cite{Cheng2008}. In Figs.~\ref{fig:spreadlow}--\ref{fig:sigma}, the markers correspond to simulation results and it can be seen that the approximations $r\ll2\rho$ and $\smin\approx\smax$ made in Theorem~\ref{thm:qj} are reasonable and that the theoretical values of the weighted communication rate accurately predict the simulated data.

Fig.~\ref{fig:spreadlow} shows the downlink rate in \eqref{eq:f} versus the code length $n$, normalized by the total number of mobile devices in the area $M$, for $M = 500$. Note that, for
$$
	\beta = \frac{\bbar M}{n} > N,
$$
a device has the capacity to cache more than the entire library, which is inefficient since there are no more files to cache. Hence, we consider only $n/M \ge \bbar/N = 10^{-2}$. The optimal content allocation $\ba^*$ is obtained by solving \eqref{P2} using a branch-and-bound method with a guarantee to have attained the best bound, i.e., the optimality gap goes to zero. Also included in Fig.~\ref{fig:spreadlow} is the downlink rate when using the strict content allocation resulting from Algorithm~\ref{alg:strict} (dashdotted black curve) with the optimal content allocation as input for an overhead $\delta = 0$. We observe that there is only a small difference in the downlink rate when using the optimal content allocation and the integer-relaxed content allocation for all values $n/M$. Note that $n/M = 1$ corresponds to maximal spreading \cite{Leong2012}, i.e., storing a coded packet of a given file on as many devices as possible. We observe that the downlink rate decreases as $n$ increases for the integer-relaxed optimal content allocation, i.e., maximal spreading \cite{Leong2012} appears to be optimal. In Fig.~\ref{fig:spreadlow}, for $n=M$,
$$
	h(\arlx) \approx h(\ba^*) \approx h(\afloor),
$$
which implies that Algorithm~\ref{alg:round} has not modified much the integer-relaxed optimal content allocation found by solving the LP \eqref{P3}, i.e., the integer-relaxed content allocation was already providing close-to-optimal and valid $\alpha_i$. For the popular content allocation, there is a tradeoff between a small $n/M$ (large $\beta$), i.e., a smaller fraction of the mobile devices cache more files from the library, and a large $n/M$. Interestingly, using the strict content allocation with $\delta = 0$ does not incur a big loss as compared to when using the optimal content allocation. In fact, for $\delta = 0.1$ (not shown in the figure), it is impossible to distinguish the downlink rate curves when using the optimal and strict content allocations. We observe from the figure that using the optimal content allocation incurs a reduction of roughly 18\% as compared to when using the popular content allocation.

Fig.~\ref{fig:spreadhigh} shows the weighted communication rate versus the code length $n$, normalized by the number of devices $M$, for $M = 2000$ and $\theta = 0.75$. Note that for such a large $M$, and consequently large $n$ when $n = M$, solving the MILP \eqref{P2} is not feasible, and instead the LP \eqref{P3} is solved. We observe that $h(\arlx) \approx h(\afloor)$ for $n = M$, i.e., both are good approximations of $h(\ba^*)$, using \eqref{eq:sandwitch}, and that $h(\arlx)$ decreases with $n$. The corresponding reduction resulting from using the optimal content allocation instead of the popular content allocation is around 25\%.

In the subsequent figures, we only consider maximal spreading, i.e., $n = M$, for the optimal content allocation and exhaustively search for the optimal $n$ when using the popular content allocation.
Fig.~\ref{fig:density} shows the weighted communication rate in \eqref{eq:h} versus the density of mobile devices in the area $M/A$ using the various content allocations. For comparison purposes, the weighted communication rate when there is no caching, $\anc = \0_N$, is also included in the figure. Using \eqref{eq:hexpand}, it is trivial to obtain
$$
	h(\anc) = M\omega\theta.
$$
We first note that the round-to-integer content allocation achieves a weighted communication rate very close to that of the integer-relaxed optimal content allocation lower bound. The infliction point observed for the optimal allocation ($M/A \approx 0.35$) corresponds roughly to the value of $M$ for which the expected aggregate cache capacity of caching devices within range of each device exceeds the number of files in the library. Note that using \eqref{eq:qj} with $n=M$ gives that the expected number of devices within range is $\lambda/\mu$ which is given by \eqref{eq:lambda} and \eqref{eq:mu}. Since each device has the capacity to cache $\bbar$ files, the expected aggregate cache capacity of caching devices within range is $\bbar \lambda/\mu$ and
\begin{align*}
	& \bbar \frac{\lambda}{\mu} = \bbar \frac{\pi}{\pi} \frac{\lambda}{\mu} = \bbar (M-1) \frac{\pi r^2}{A} > N\\
	\Longrightarrow \quad & \frac{M}{A} > \frac{N}{\bbar} \cdot \frac{1}{\pi r^2} + \frac{1}{A} \approx \frac{N}{\bbar} \cdot \frac{1}{\pi r^2} \approx 0.32,
\end{align*}
where the first approximation holds for a large area $A$. We see that, using the optimal content allocation, we can effectively leverage the available cache size and reduce the weighted communication rate by around $38\%$ compared to when using the popular content allocation. For a larger density of devices ($M/A\approx0.8$), the difference in weighted communication rate is reduced since the expected aggregate cache capacity within range of any reference device is very large. In this case, a device requesting a particular file is likely to find coded packets cached by devices within the communication range also when using the popular content allocation. Despite this fact, the reduction in the weighted communication rate of using the optimal content allocation instead of the popular content allocation is still around $18\%$.

In Fig.~\ref{fig:weight}, the weighted communication rate is plotted versus the weighting parameter $\theta$. As explained by Proposition~\ref{prop:popopt}, the popular content allocation is optimal for $\theta = 0.5$. The gain over no caching, i.e., all files are downloaded from the BS, observed for $\theta = 0.5$ is due to devices self-servicing, i.e., finding requested content in the own cache. For the popular content allocation and $\theta < 0.54$, maximal spreading, i.e., a smaller number of files cached by all devices, entails a lower weighted communication rate. For larger $\theta$, a reduced spreading is desirable. For $M = 2000$ and $0.54 \le \theta < 0.58$, $n/M = 0.04$ is found to be optimal. For $\theta \ge 0.58$, an exhaustive search reveals that $n/M = 0.025$ is optimal. We also see that the reduction in the weighted communication rate entailed by the optimal content allocation instead of the popular content allocation increases with $\theta$.

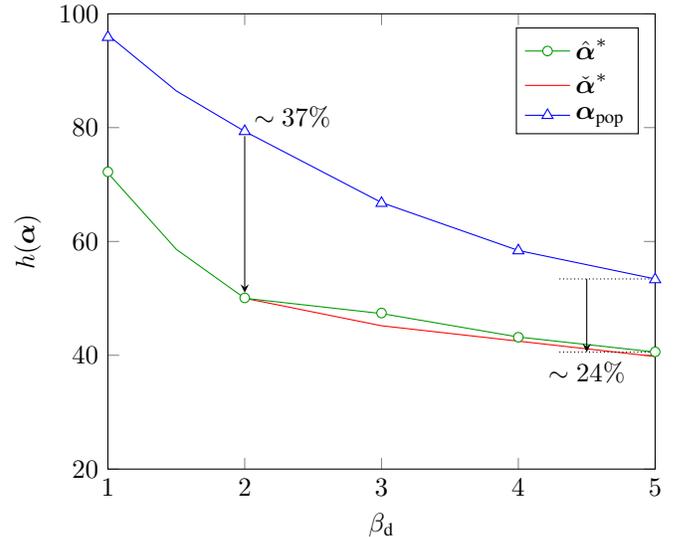
\begin{figure}[t!]
	\centering
	\begin{tikzpicture}[>=stealth]
\begin{axis}[%
	width = \columnwidth,
	xmin = 1,
	xmax = 5,
	xlabel = {$\bbar$},
	ymin = 20,
	ymax = 100,
	ylabel = {$h(\ba)$},
	legend cell align = left,
	legend pos = north east,
]

\addplot [color = color1, solid, forget plot]
	table[row sep=crcr]{%
		2	50.01918\\
		3	45.18512\\
		5	39.77968\\
};
\addplot [color = color2, solid, forget plot]
	table[row sep=crcr]{%
		1	72.01789\\
		1.5	58.6167\\
		2	50.0204\\
		3	47.3334\\
		4	43.2120\\
		5	40.5624\\
	};
\addplot [color = color3, solid, forget plot]
	table[row sep=crcr]{%
		1	96.2873\\
		1.5	86.4495\\
		2	79.3420\\
		3	66.8362\\
		4	58.4149\\
		5	53.4150\\
	};

% Simulations >>>>>>>>>>>>>>>>>>>>>>>
\addplot [color = color3, only marks, every mark/.append style = {fill = white, scale = 1.2}, mark = triangle*, forget plot]
	table[row sep=crcr]{%
		1	95.885\\
		2	79.342\\
		3	66.72356\\
		4	58.3874\\
		5	53.3067\\
	};

\addplot [color=color2, only marks, every mark/.append style = {fill = white, scale = .9}, mark = *, forget plot]
	table[row sep=crcr]{%
		1	72.23162\\
		2	50.0769\\
		3	47.4235\\
		4	43.1669\\
		5	40.6013\\
	};
\draw[<-] ($(axis cs:2, 51)$) to node[above, at end, anchor = south west]{$\sim37\%$} ($(axis cs:2, 78.5)$);
\draw[densely dotted] (axis cs:4.3, 53.41498) to (axis cs:5, 53.41498);
\draw[densely dotted] (axis cs:4.3, 40.5624) to (axis cs:5, 40.5624);
\draw[<-] (axis cs:4.5, 40.5624) to node[below, at start, anchor = north]{$\sim24\%$} (axis cs:4.5, 53.41498);

\addlegendimage{color = color2, every mark/.append style = {fill = white, scale = 0.9}, mark = *};
\addlegendentry{$\afloor$};
\addlegendimage{color = color1};
\addlegendentry{$\arlx$};
\addlegendimage{color = color3, every mark/.append style = {fill = white, scale = 1.2}, mark = triangle*};
\addlegendentry{$\apop$};
%\addlegendimage{color = color3, every mark/.append style = {fill = white, scale = .9}, mark = *};
\end{axis}
\end{tikzpicture}%
	\vspace{-3ex}
	\caption{The weighted communication rate $h(\ba)$ versus the average cache size constraint per device $\bbar$ using different content allocations for $M = 2000$, $\sigma = 0.7$, $\theta = 0.75$. All markers correspond to simulated rate.}
	\label{fig:bbar}
	\vspace{-3ex}
\end{figure}
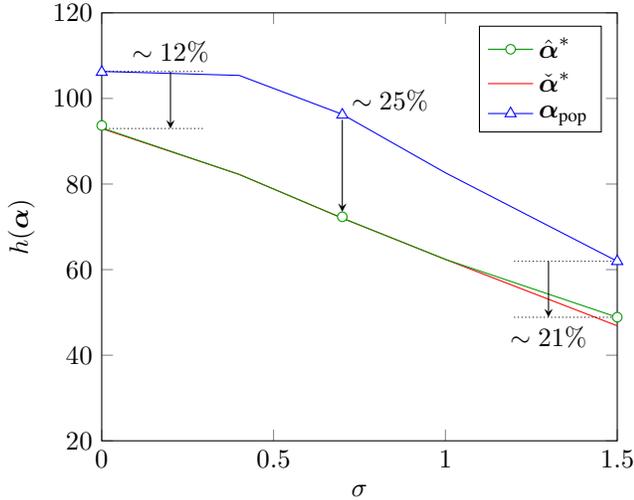
\begin{figure}[t!]
	\centering
\begin{tikzpicture}[>=stealth]

%\pgfplotsset{set layers}
%
%\begin{axis}[
%	xmin = 0,
%	xmax = 2,
%	ymin = 0,
%	ymax = 40,
%	axis y line* = right,
%	axis x line = none,
%	ylabel = {Gain [\%]}]
%	\addplot [color = black, densely dotted, forget plot]
%		table[row sep=crcr]{%
%			0	12.55\\
%			0.4	21.66\\
%			0.7	24.90\\
%			1	24.22\\
%			1.5	23.81\\
%			2	31.15\\
%	};
%\end{axis}
\begin{axis}[%
	xmin = 0,
	xmax = 1.5,
	xlabel = {$\sigma$},
	xtick = {0, 0.5, ..., 1.5},
	ymin = 20,
	ymax = 120,
%	axis y line* = left,
	ylabel = {$h(\ba)$},
	legend cell align = left,
	legend pos = north east,
]

%\draw[dashed] (axis cs:0.6, 26.1) to (axis cs:0.002, 26.1);
%\draw[dashed] (axis cs:0.02, 33.4983) to (axis cs:0.002, 33.4983);
%\draw[dashed] (axis cs:0.01, 50) to (axis cs:0.6, 50);

\addplot [color = color1, solid, forget plot]
	table[row sep=crcr]{%
		0	92.97223\\
		0.4	82.24592\\
		0.7	72.02576\\
		1	62.38826\\
		1.5	46.86929\\
%		2	29.9679\\
};
\addplot [color = color2, solid, forget plot]
	table[row sep=crcr]{%
		0	93.133\\
		0.4	82.2468\\
		0.7	72.0273\\
		1	62.3893\\
		1.5	48.8698\\
%		2	40.0963\\
};
\addplot [color = color3, solid, forget plot]
	table[row sep=crcr]{%
		0	106.3174\\
		0.4	105.3838\\
		0.7	96.2873\\
		1	82.6278\\
		1.5	61.9514\\
%		2	43.5287\\
	};

% Simulations >>>>>>>>>>>>>>>>>>>>>>>
\addplot [color = color3, only marks, every mark/.append style = {fill = white, scale = 1.2}, mark = triangle*, forget plot]
	table[row sep=crcr]{%
		0	106.1417\\
		0.7	96.1519\\
		1.5	61.9318\\
	};
\addplot [color = color2, only marks, every mark/.append style = {fill = white, scale = 0.9}, mark = *, forget plot]
	table[row sep=crcr]{%
		0	93.681\\
		0.7	72.3421\\
		1.5	48.8899\\
	};
%
%\addplot [color=color1, only marks, every mark/.append style = {fill = white, scale = .8}, mark = square*, forget plot]
%	table[row sep=crcr]{%
%		0	93.49622\\
%		0.7	72.23162\\
%		1.5	46.82781\\
%	};

\draw[densely dotted] (axis cs:0, 106.3174) to (axis cs:0.3, 106.3174);
\draw[densely dotted] (axis cs:0, 92.97223) to (axis cs:0.3, 92.97223);
\draw[<-] (axis cs:0.2, 92.97223) to node[above, at end, anchor = south]{$\sim12\%$} (axis cs:0.2, 106.3174);
\draw[<-] (axis cs:0.7, 73.5) to node[above, at end, anchor = south west]{$\sim25\%$} (axis cs:0.7, 95);
\draw[densely dotted] (axis cs:1.2, 61.9514) to (axis cs:1.5, 61.9514);
\draw[densely dotted] (axis cs:1.2, 48.8698) to (axis cs:1.5, 48.8698);
\draw[<-] (axis cs:1.3, 48.8698) to node[below, at start, anchor = north]{$\sim21\%$} (axis cs:1.3, 61.9514);

	\addlegendimage{color = color2, every mark/.append style = {fill = white, scale = 0.9}, mark = *};
	\addlegendentry{$\afloor$};
	\addlegendimage{color = color1};
	\addlegendentry{$\arlx$};
	\addlegendimage{color = color3, every mark/.append style = {fill = white, scale = 1.2}, mark = triangle*};
	\addlegendentry{$\apop$};
\end{axis}
\end{tikzpicture}%
	\vspace{-3ex}
	\caption{The weighted communication rate versus the Zipf parameter $\sigma$ using different content allocations for $M = 2000$, $\theta = 0.75$, and $\bbar = 1$. Markers correspond to simulated rate.}
	\label{fig:sigma}
	\vspace{-3ex}
\end{figure}

Fig.~\ref{fig:bbar} shows the weighted communication rate versus the average cache size constraint per device $\bbar$. Recall that, for the optimal content allocation with maximal spreading, the cache constraint is strict. We see that using the optimal content allocation instead of the popular content allocation entails a significant reduction in the weighted communication rate for some $\bbar$. As $\bbar \to N$ the reduction vanishes, which is intuitive as each device can cache the entire file library and selecting a content allocation is no longer relevant.

Finally, in Fig.~\ref{fig:sigma}, the weighted communication rate is plotted versus the Zipf parameter $\sigma$. Note that, for $\sigma \rightarrow 0$, the file popularity distribution approaches the uniform distribution. For $\sigma=0$, as expected, uniform content allocation, i.e., $\alpha_i=\alpha$ $\forall i$, is optimal. In other words, all files are cached using the same $(n,k)$ MDS code. Using the round-to-integer content allocation instead of the popular content allocation leads to a reduction of the weighted communication rate of around $12\%$, which is due to the more efficient use of the cache space when using the former allocation, i.e., the probability of redundant content being cached in devices within the communication range is negligible. For larger $\sigma$ the distribution is more skewed towards the most popular files and the weighted communication rate decreases. The reason is that less files have a notable popularity, less files are frequently requested, and the fixed cache size constraint allows these few popular files to be cached. In this case the optimal content allocation deviates from the uniform one. For $\sigma=1.5$, the weighted communication rate is decreased by approximately $21\%$ when the round-to-integer content allocation is used instead of the popular content allocation. This is because the round-to-integer content allocation uses some of the available cache space to cache coded packets from the tail of less frequently requested files that cumulatively adds up to a non-negligible fraction of the requests.

\section{Conclusion}
We optimized the caching of content in mobile devices using maximum distance separable codes. We derived a good approximation of the distribution of the number of caching devices within range of a device as the devices move around according to the random waypoint model. We formulated a mixed integer linear program to minimize the weighted sum of the downlink rate and the device-to-device communication rate under a global average cache size constraint. We showed that optimized MDS coded caching yields a significantly lower weighted communication rate compared to when caching (uncoded) the most popular files, especially when the device density is high. Furthermore, we showed numerically that caching coded packets of a particular file on all devices, i.e., maximal spreading, is optimal.

\bibliographystyle{IEEEtran}
%\bibliography{IEEEabrv,confs-jrnls,library}

\end{document}